# Transport and pinning properties of Ag-doped $FeSe_{0.94}$

E. Nazarova[1,2], N. Balchev[1], K. Nenkov[2,4], K. Buchkov[1], D. Kovacheva[3], A. Zahariev[1] and G. Fuchs[2]

[1]*Institute of Solid State Physics, Bulgarian Academy of Sciences, 1784 Sofia, Bulgaria*
[2]*Leibniz Institute for Solid State and Materials Research, (IFW Dresden), P.O. Box 2700116, D-01171 Dresden, Germany*
[3]*Institute of General and Inorganic Chemistry, Bulgarian Academy of Sciences, 1113 Sofia, Bulgaria*
[4]*International Laboratory of High Magnetic Fields and Low Temperatures, Gajowicka 95, 53-529 Wroclaw, Poland*

E-mail: nbalchev@yahoo.com

**Abstract** We investigated the superconducting transition and the pinning properties of undoped and Ag-doped $FeSe_{0.94}$ at magnetic fields up to 14 T. It was established that due to Ag addition the hexagonal phase formation in melted $FeSe_{0.94}$ samples is suppressed and the grain connectivity is strongly improved. The obtained superconducting zero-field transition becomes sharp (with a transition width below 1 K), $T_c$ and the upper critical field were found to increase, whereas the normal state resistivity significantly reduces becoming comparable with those of FeSe single crystals. In addition, a considerable magnetoresistance was observed due to Ag doping. The resistive transition of undoped and Ag-doped $FeSe_{0.94}$ is dominated by thermally activated flux flow. From the activation energy U vs H dependence, a crossover from single-vortex pinning to a collective creep pinning behavior was found with increasing the magnetic field.

## 1. Introduction

Since the discovery of superconductivity in tetragonal FeSe phase [1] this compound is an intensively investigated iron based superconductor. Up to now different families of superconductors based on Fe layers are discovered. The most studied are called ‘‘1111’’ for REFeAsO, ‘‘122’’ for $AFe_2As_2$, ‘‘111’’ for AFeAs and ‘‘11’’ for Fe(Te,Se) [2,3]. All these families have FeAs or FeSe planes as their common building blocks, which are responsible for the superconductivity [4]. The 11 type materials have a rather simple crystal structure without the charge reservoir layer and with only two elements [5]. These materials are of interest both for the technological applications and for the understanding of the vortex properties in the mixed state due to their simple structure and nearly isotropic upper critical field. Superconductivity in the tetragonal β-FeSe phase was related to a selenium deficient $FeSe_{0.92}$ composition [6]. But the β-FeSe phase which was reported to be extremely sensitive to stoichiometry and preparation conditions was later realized in a sample of $Fe_{1.01}Se$ ($FeSe_{0.99}$) composition prepared without spurious oxides and oxygen defects [7]. The phase diagram of FeSe is very complex [8], and other non-superconducting hexagonal phases, as the α-FeSe phase (of NiAs structure) and the α-$Fe_7Se_8$ phase can appear by preparing β-FeSe. Flux methods are used to prepare FeSe based single crystals. In a second step, β-FeSe crystals are separated from the flux. Their superconducting transition temperature is, with $T_c$ ~ 9.5 - 10 K (midpoint value), usually at least by 1 K higher than that of polycrystalline β-FeSe samples. The iron-based superconductors exhibit rich vortex phenomena in the mixed state due to their large upper critical fields and small lower critical fields [5]. The vortex dynamic of the 11 system was recently studied by investigating $\beta$-FeSe single crystals [9].

As already shown in other high temperature superconductors (see for example [10,11]), silver has been widely used as a dopant or additive to improve the microstructure and superconducting (SC) properties. The role of Ag addition on structure and SC properties was investigated for polycrystalline $Sr_{0.6}K_{0.4}Fe_2As_2$ [12] and $FeSe_{0.5}Te_{0.5}$ [13]. It was established that silver improves the intergrain connections and enhances the critical current $J_c$ of these superconductors. Moreover, a small amount of Ag was found to enter into the crystal structure of $FeSe_{0.5}Te_{0.5}$. In our previous work [14] we established that a small amount of Ag incorporated in the grains of $FeSe_{0.94}$ improves both intra- and inter-granular superconducting properties. It was also found that due to Ag doping the irreversibility field slightly increases and the intergrain connections improve. The purpose of this work is to investigate the resistive transition and pinning properties of Ag-doped polycrystalline $FeSe_{0.94}$ in the presence of magnetic fields.

## 2. Experiments

The investigated samples with nominal compositions $FeSe_{0.94}$ and $FeSe_{0.94}$ +4wt % Ag are obtained by partial melting. The initial products Se, Fe and Ag powders with purity 99,9%, 99,5% and 99,9% respectively are mixed and pressed into tablets in a glove box with Ar atmosphere. The tablets are put in silica tube, evacuated to $10^{-4}$ torr and sealed. The heat treatment is performed in vacuum furnace at $700^0$C for 8 hours. After a new grinding and pressing, the tablets are sealed in double evacuated quartz tubes for the melting process. The heat treatment is performed at $1050^0$C for 20 h. The synthesis is completed by an additional annealing at $400^0$C for 100 h followed by furnace cooling. Undoped $FeSe_{0.94}$ is also obtained by the solid state reaction (SSR) method. In this case the second synthesis is provided at $700^0$C for 24 h in single evacuated quartz tube and the samples are also annealed at $400^0$C for 36 h followed by furnace cooling.

Powder X-ray diffraction patterns are collected within the range from 5.3 to 80° 2θ with a constant step 0.02° 2θ on Bruker D8 Advance diffractometer with Cu Kα radiation and LynxEye detector. Phase identification is performed with the Diffrac*plus* EVA using ICDD-PDF2 Database. The resistivity of the samples at different magnetic fields is measured by the standard four probe method using DC resistivity option of PPMS-14 T. The external DC magnetic field is always perpendicular to the sample (a,b) plane.

## 3. Results and discussion

In Fig. 1 the XRD patterns of the investigated samples are shown. According to these diffractograms, both pure samples consist mainly of the tetragonal phase. However, traces of a hexagonal phase are visible in the sample prepared by melting. In contrast, no hexagonal phase is detected in the sample with Ag addition. Only, a small amount of Ag impurities was found. Thus, analogous to the Sn addition in FeSe [15], Ag seems to help to suppress the formation of the hexagonal phase increasing in this way the tetragonal phase fraction. The lattice parameters in both samples prepared by melting (pure and with Ag addition) are almost the same (a=3.7650 Ǻ; c=5.5180 Ǻ and a=3.7671 Ǻ; c=5.5193 Ǻ respectively) indicating that Ag is probably not incorporated in the unit cell. SEM observations show a non-homogeneous Ag distribution. Two types of grains have been detected by EDX analysis: rarely occurring grains with Ag content of about 90% and regular FeSe grains with silver content of only few percent. Upon lowering the accelerating voltage, the resulting signal comes from the thinner surface layer. In this case, the silver content in the regular grains increases about 2 times. This suggests that Ag is distributed mainly on the grain surface boundaries.

In Fig. 2, the resistivity vs. temperature dependence is shown for both undoped and the Ag doped sample at H=0. A characteristic bump in the normal state resistivity is observed at around 90 K which is related to a structural phase transition from tetragonal at room temperature to orthorhombic at ~ 80 K in the investigated samples. This structural transition was detected soon after the discovery of superconductivity in $FeSe_{1-x}$ [6]. In some of iron based compounds, as 1111 [16], 122 [17] and 111 [18] compounds, the structural transition correlates with the appearance of magnetic order probably responsible for electron pairing and superconductivity. Density functional studies of FeSe also show a spin density wave ground state [19]. Experimentally, a specific change in FC magnetic susceptibility is observed starting around 100 K and finishing at 75-80 K [20] which is believed to arise from immobile interacting magnetic defects. We find the same behavior for the undoped sample as shown in Fig. 3 where dρ/dT and the magnetic moment are plotted against the temperature. Both dependences show anomalies at ~ 80 K, connected with the structural phase transition and magnetic changes in the above mentioned temperature interval.

Worth mentioning is the almost linear temperature dependence of the resistivity in normal state for all samples for the orthorhombic phase at temperatures between $T_c$ and ~ 90 K. This behavior is similar to the cuprates above $T_c$. Both types of high temperature superconductors (Fe-based and cuprates) show non-Fermi liquid behavior above the superconducting transition. For Fermi liquids, the electron-electron scattering is dominating and the resistance varies as $T^2$ at low temperatures.

As shown in Fig. 2, the undoped $FeSe_{0.94}$ sample obtained by melting has a higher resistivity in the normal state than the sample prepared by SSR. This might be associated with the presence of the hexagonal phase. However, a very low normal state resistivity is achieved by Ag doping suggesting an almost single-phase sample without hexagonal phase (in accordance with XRD results).

Superconductivity appears at temperatures below ~9K. Remarkably, the critical temperature $T_c^{50}$ (at the midpoint of the transition) of the Ag doped $FeSe_{0.94}$ sample is shifted to 9.4 K which is ~ 1 K higher than $T_c^{50}$ of the undoped samples (see inset of Fig. 2). Also the temperature $T_{c(R=0)}$ at which the resistivity completely disappears increases for the Ag doped sample to 9.0 K compared to $T_{c(R=0}$ ~ 7.5 K for both undoped $FeSe_{0.94}$ samples which is a typical value for undoped polycrystalline selenium-deficient $FeSe_{1-x}$ [21].

In Fig. 4, the superconducting transition of the Ag doped sample is compared with the transition curve reported for a $\beta$-FeSe single crystal [9]. It is clearly seen that the Ag doped sample has a sharper transition curve than the single crystal which correlates with its low normal state resistivity $\rho_N$. This low $\rho_N$ ~ 0.3 mΩ cm of the Ag doped sample is probably related to the FeSe grains and can be attributed to impurity scattering in FeSe assuming that the superconductor is in the dirty limit. Lower values of $\rho_N$ ~ 0.2 mΩ cm [8] and ~ 0.1 mΩ cm [22] have been reported for FeSe single crystals suggesting a larger mean free path of the charge carriers than in the Ag doped sample.

In Table 1, the superconducting zero-field properties of the Ag doped $FeSe_{0.94}$ sample are compared with data obtained for the $\beta$-FeSe single crystal [9]. It can be concluded that due to Ag doping of polycrystalline $FeSe_{0.94}$, a significant improvement of the zero-field properties can be achieved. The excellent quality of the Ag doped sample can be explained by the disappearance of the non-superconducting hexagonal phase and other inhomogeneities, but also by the strongly improved intergrain connections of this polycrystalline sample.

The upper critical field $B_{c2}$ is an important parameter both from the fundamental and practical point of view. It gives information on the pair-breaking mechanisms in magnetic field and for superconducting parameters like coherence length and anisotropic parameter. We performed measurements of resistance vs. temperature at different magnetic fields up to 14 Tesla in order to determine the temperature dependent $B_{c2}(T)$. In Fig. 5, $\rho(T)$ data are presented for the undoped and the Ag doped sample prepared by melting. The shift of the transition curves to lower temperatures for increasing magnetic field is accompanied by a broadening of the transition curves which is more pronounced for the undoped sample.

Unexpected is the strongly enhanced magneto-resistance (MR) observed in the Ag doped sample. In Fig. 6, MR = $\Delta\rho/\rho_o$ [with $\Delta\rho = (\rho(T,B) - \rho_o)$ and $\rho_o=\rho(T,0)$] of the Ag doped sample is plotted against the magnetic field at different temperatures. The MR data can be described in the form of the Kohler plot, i.e. the $\Delta\rho/\rho_o$ data plotted against [$B\cdot\rho(300\text{ K})/\rho_o$] fall on a single line (see inset of Fig. 5). This is very surprising because the Kohler scaling is believed to be not compatible with non-Fermi liquid behaviour which we observed in our samples. Further investigations are required in order to resolve this puzzle, which is, however, beyond the subject of our paper. The almost linear field dependence of the MR observed in the Ag doped sample suggests the possible formation of silver chalcogenides due to Ag doping of the $FeSe_{0.94}$ sample. For $Ag_{2+\delta}Se$ and $Ag_{0.7}Se_{0.7}$, a rather large MR of ~ 3 and ~ 1 (at 10 K and 5T), respectively, was reported recently [23] which was found to increase almost linear with the applied magnetic field without tendency of saturation up to at least 5.5 T. Another possible scenario might be that the observed MR is related to the presence of Ag particles which exhibit a significant MR, too [24].

In Fig. 7, $B_{c2}(T)$ data are compared using the midpoint of the resistive transition ($0.5\rho_n$) in Fig. 4 to define the upper critical field $B_{c2}$. According to the standard Werthamer-Helfand-Hohenberg (WHH) model [25] for type II superconductors, the orbital upper critical field $B_{c2}^*(0)$ at $T = 0$ is given by

$$B_{c2}^*(0) = -0.69\, T_c\, (\mathrm{d}B_{c2}/\mathrm{d}T)|_{T=T_c}\,, \qquad (1)$$

where $(\mathrm{d}B_{c2}/\mathrm{d}T)|_{T=T_c}$ is the slope of $B_{c2}(T)$ at $T_c$. The upper critical field $B_{c2}$ for the Ag doped $FeSe_{0.94}$ sample is found to increase with a similar slope of -4.4 T/K by lowering the temperature as found for the undoped molten sample. Therefore, the high $B_{c2}(0) \sim 28.1$ T at $T = 0$ of the Ag doped sample is mainly due to its enhanced $T_c$. In contrast, a low $B_{c2}(0) \sim 17.0$ T is obtained by SSR which is caused both by a lower $T_c \sim 8.3$ K and a smaller slope of $B_{c2}(T)$.

The experimental data in Fig. 7 can be described in a restricted field range by the orbital upper critical field $B_{c2}^*(T)$ of the WHH model [25] neglecting spin-paramagnetic effects and ignoring spin-orbit scattering. The fit curves are shown in Fig. 7 by dotted lines. By fitting the data, the pronounced positive curvature of the $B_{c2}$ data near $T_c$ was neglected. Furthermore, it is seen in Fig. 6 that the experimental data can be described by the WHH model only up to fields of about 5 T, whereas at higher fields the data exceed the orbital $B_{c2}$. Both deviations from the single-band WHH model are clear indications for multi-band superconductivity in the investigated $FeSe_{0.94}$ samples. Multi-band superconductivity has been reported also for $\beta$-FeSe single crystals. The measured $B_{c2}(T)$ of ~ 17.5 T at $T$ ~ 0.3 K for a single crystal was found to exceed the orbital field of $B_{c2}^*(0)$ = 15.3 T significantly which was explained by multi-band superconductivity [26].

From the obtained $B_{c2}^*(0)$ data, the Ginzburg-Landau (GL) coherence length ξ can be estimated using the relation

$$B_{c2}(0) = \Phi_0/2\pi\,\xi(0)^2 \qquad (2)$$

neglecting the multi-band superconductivity in $FeSe_{0.94}$. In Eq. (2), $\Phi_0$ = $2.068\cdot10^{-15}$ T·m$^2$ is the flux quantum. The obtained coherence lengths are given in Table 1, where also other superconducting parameters of the investigated samples are presented. The coherence length of $FeSe_{0.94}$ is, with ξ(0) ~ 10 $a$, much larger than the unit cell $a$ suggesting that no weak-link behavior is expected in $FeSe_{0.94}$. In addition, $FeSe_{0.94}$ has only a small anisotropy. Taking this and the absence of weak links into account, would explain why polycrystalline single-phase $FeSe_{0.94}$ samples, such as the investigated Ag doped sample, can have similar or even better properties than single crystals. It should be also noted that the estimated GL coherence lengths are much smaller than the BCS coherence length $\xi_o = 0.15\ h v_F/(2\pi\ k_B T_c)$ ~ 7.9 nm. Here, the Fermi velocity $v_F$ ~ 0.4 eVÅ reported for $Fe_{1.03}Te_{0.7}Se_{0.3}$ [27] was used. The relation $\xi(0) < \xi_o$ confirms our assumption that $l < \xi_o$ (with $l$ as the mean free path of the charge carriers), i.e. the investigated $FeSe_{0.94}$ samples are in the dirty limit.

The lower critical field $H_{c1}$ for the Ag doped sample was estimated from magnetization data as shown in Fig. 8. At 2 K, a diamagnetic signal is observed for fields in the range up to 240 Oe. The estimated $H_{c1}$(2K) ~ 30 Oe in the Meissner state is in reasonable agreement with the precisely determined lower critical field in FeSe at 2 K [28]. $H_{c1}$ is found to increase slightly to 31.7 Oe at $T$ = 0 using the expression $H_{c1}(T)=H_{c1}(0)[1-(T/T_c)^2]$. From the GL equation, $H_{c1}(0)=[\Phi_0/4\pi\lambda(0)^2]\ln k$, where λ(0) is the London penetration depth and κ=λ(0)/ξ(0) is the GL parameter, λ(0) ≈ 326 nm and κ~104 were derived using the already determined values for $H_{c1}(0)$ and ξ(0). The high GL parameter κ indicates that the Ag doped sample is a so-called extreme type II superconductor.

The resistive superconducting transition in $FeSe_{0.94}$ is strongly influenced by its flux dynamic. The increasing broadening of the transition curves in magnetic fields (see Fig. 6) is related to thermally activated flux flow (TAFF) [29]. The resistivity in the TAFF region is given by

$$\rho(H,T) = \rho_{of}\exp(-U(T,H)/k_B T) \qquad (3)$$

with $U(T,H)$ as the activation energy, and the prefactor $\rho_{of}$, which is assumed to be constant similar as for high-$T_c$ superconductors. Using the expression $U(T,H)$

$= U_o(H)(1-T/T_c)$ for the activiation energy, one gets from Eq. (3) the Arrhenius relation:

$$\ln \rho(H,T) = \ln \rho_O(H) - U_o(H)/k_B T \quad (4)$$

where $\rho_O(H)$ is given by

$$\ln \rho_O(H) = \ln \rho_{of} + U_o(H)/T_c \quad (5)$$

The Arrhenius plots (see Eq. (4) for the undoped and Ag doped molten $FeSe_{0.94}$ samples are shown in Figure 9. In the TAFF region, the ln ρ versus $1/T$ data can be described by straight lines. The slopes of the straight lines and its y intercept for $1/T = 0$ correspond to $U_o(H)$ and $\rho_O(H)$, respectively. The $\rho(T,H)$ data have a common $T_c$ which is given by the temperature $T_c^{cross}$ of the crossing point of the straight lines. $T_c$ can be also determined by plotting of ln $\rho_O(H)$ against $U_o(H)$ (see Fig. 10). According to Eq. (5), the slopes of the straight lines in Fig. 10 amount to $1/T_c$.

As shown in Table 2, the obtained $T_c$ values are found to be consistent with $T_c^{cross}$ derived form Fig. 9. $T_c$ (or $T_c^{cross}$) exceeds $T_c^{50}$ (defined at the midpoint of the transition curves) by ~ 0.2 K for the undoped sample and by ~ 0.3 K for the Ag-doped sample.

In Fig. 11, the field dependence of the activation energy, $U_o(H)$, is compared for the undoped and Ag doped molten samples. The activation energy shows a weak field dependence at low fields changing in a stronger field dependence at $\mu_o H$ ~ 4T. Both branches can be described by a power law behavior, $U_o \sim H^{-\alpha}$, as indicated by the straight lines in this double-logarithmic plot. The effect of Ag doping on the activation energy is two-fold: 1) $U_o$ increases by a factor of ~ 1.5 in both branches due to Ag doping and 2) additionally $U_o$ strongly enhances at low fields ($\mu_o H < 4$T) due to Ag doping corresponding to an increase of α from 0.15 for the undoped sample to 0.57 for the Ag doped sample. At high fields ($\mu_o H > 4$T), the field dependence of $U_o$ becomes strong for both samples, i.e. α is found to increase to α = 1.65 for both samples. Nevertheless, $U_o$ of the Ag doped sample is 50% higher than for the undoped sample as mentioned above.

A similar field dependence of $U_o(H)$ was also reported for β-FeSe [9] and Fe(Te,S) single crystals [30] and associated with a crossover from single-vortex pinning at low magnetic fields to collective pinning at high fields. For comparison, the data for the β-FeSe single crystal are included in Fig. 11. These data perfectly agree with those of the Ag doped sample above ~ 4T, whereas the $U_o(H)$ data for fields < 4T remain below $U_o(H)$ of the Ag doped sample. In the single-vortex pinning regime the correlation length remains less than the inter-vortex spacing $a_0 \sim (\Phi_0/B)^{1/2}$, i.e. the vortices are pinned independently so that $j_c$ and the activation energy $U_o$ is almost field-independent. At high enough fields the inter-vortex interaction becomes significant and $U_o$ is controlled by collective pinning of vortex bundles confined by a field dependent correlation volume [31]. The crossover to collective pinning observed in Fig. 11 is found to occur at $a_0$ = 24.4 nm.

According to Fig. 11, by Ag doping the activation energy for thermally activated flux flow strongly enhances at low fields suggesting improved flux pinning in that sample. This high activation energy is responsible for the sharp transition curves found for the Ag doped sample (see, Figs. 4 and 5b) which contrasts with the broad transition curves reported for a $\beta$-FeSe single crystal [9] (see also Fig. 4) which can be attributed to its relatively low activation energy.

## 3. Conclusion

Investigating the superconducting transition of undoped and Ag doped $FeSe_{0.94}$ by resistance measurements in magnetic fields up to 14 T it was found that due to Ag addition the undesired hexagonal phase formation is suppressed, and the superconducting properties significantly improved. In particular, the superconducting transition temperature increased by ~ 1K on 9.4 K (midpoint of transition) and the superconducting transition, with transition widths below ~ 1K, becomes sharper than those of $\beta$-FeSe single crystals. The sharp superconducting transition of the Ag doped sample is closely related to its high activation energy for thermally activated flux flow which is dominating a large portion of the resistive transition. It was also found that Ag doping enhances the magnetoresistance and the upper critical field up to ~ 28 T which was obtained by extrapolating the experimental data to $T$ = 0 by using the standard WHH model for the orbital $B_{c2}{}^*(T)$.

## Acknowledgement

This work was supported by the European Atomic Energy Agency (EURATOM) through the Contract of Association Euratom-INRNE.BG. One of the authors (E. N.) is grateful to the Leibniz Institute for Solid State and Materials Research, Dresden (especially to Prof. B. Holzapfel) for the invitation and possibility to perform the experiments.

## References

[1] Hsu F C, Luo J Y, Yeh K W, Chen T K, Huang T W, Wu P M, Lee Y C, Huang Y L, Chu Y Y, Yan D C, and Wu M K 2008 *Proc. Natl. Acad. Sci. USA* **105** 14262

[2] Pallecchi I, Tropeano M, Lamura G, Pani M, Palombo M, Palenzona A, Putti M, 2012 *Physica C* **482** 68

[3] Putti M, Pallecchi I, Bellingeri E, Cimberle M R, Tropeano M, Ferdeghini C, Palenzona A, Tarantini C, Yamamoto A, Jiang J, Jaroszynski J, Kametani F, Abraimov D, Polyanskii A, Weiss J D, Hellstrom E E, Gurevich A, Larbalestier D C, Jin R, Sales B C, Sefat A S, McGuire M A, Mandrus D, Cheng P, Jia Y, Wen H H, Lee S and Eom C B, 2010 *Supercond. Sci. Technol.* **23** 034003

[4] Wen H H, 2012 *Rep. Prog. Phys*. **75** 112501

[5] Lei Hechang, Wang Kefeng, Hu Rongwei, Ryu Hyejin, Abeykoon Milinda, Bozin Emil S and Petrovic Cedomir, 2012 *Sci. Technol. Adv. Mater*. **13** 054305

[6] Margadonna S, Takabayashi Y, McDonald M T, Kasperkiewicz K, Mizuguchi Y, Takano Y, Fitch A N, Suard E and Prassides K, 2008 *Chem. Commun*. Issue **43** 5607

[7] T. M. McQueen, Q. Huang, V. Ksenofontov, C. Felser, Q. Xu, H. Zandbergen, Y. S. Hor, J. Allred, A. J. Williams, D. Qu, J. Checkelsky, N. P. Ong, and R. J. Cava, 2009 *Phys. Rev. B* **79** 014522

[8] Rongwei Hu, Hechang Lei, Milinda Abeykoon, Emil S. Bozin, Simon J. L. Billinge, J. B. Warren, Theo Siegrist, and C. Petrovic, 2011 *Phys. Rev. B* **83** 224502

[9] Lei Hechang, Hu Rongwei and Petrovic C, 2011 *Phys. Rev. B* **84** 014520

[10] Rani Poonam, Pal Anand, Awana V P S, 2014 *Physica C* **497** 19

[11] Mawassi R, Marhaba S, Roumie M, Awad R, Korek M, Hassan I, 2014 *J Supercond Nov Magn* **27** 1131

[12] Ding Qing-Ping, Prombood Trirat, Tsuchiya Yuji, Nakajima Yasuyuki and Tamegai Tsuyoshi, 2012 *Supercond. Sci. Technol*. **25** 035019

[13] Migita M, Takikawa Y, Sugai K, Takeda M, Uehara M, Kuramoto T, Takano Y, Mizuguchi Y, Kimishima Y, 2013 *Physica C* **484** 66

[14] Nazarova E, Buchkov K, Terzieva S, Nenkov K, Zahariev A, Kovacheva D, Balchev N and Fuchs G, arXiv:1401.6296 [cond-mat.supr-con]

[15] Chen Ning, Ma Zongqing, Liu Yongchang, Li Xiaoting, Cai Qi, Li Huijun, Yu Liming, 2014 *Journal of Alloys and Compounds* **588** 418

[16] de la Cruz C, Huang Q, Lynn J W, Li J, Ratcliff II W,. Zarestky J. L, Mook H A, Chen G F, Luo J L, Wang N L and Dai P 2008 *Nature*, **453** 899

[17] Huang Q, Qiu Y, Bao Wei, Green M A, Lynn J W, Gasparovic Y C, Wu T, Wu G, and Chen X H 2008 *Phys. Rev. Lett.* **101** 257003

[18] Li Shiliang, de la Cruz Clarina, Huang Q, Chen G F, Xia T.-L., Luo J L, Wang N L and Dai Pengcheng 2009 *Phys. Rev. B* **80** 020504(R)

[19] Subedi A, Zhang L J, Singh D J and Du M H 2008 *Phys Rev B* **78** 134514
[20] Lee K W, Pardo V and Pickett W E, arXiv:0808.1733
[21] Sudesh, Rani S, Varma G D 2013 *Physica C* **485** 137
[22] Braithwaite D, Salce B, Lapertot G, Bourdarot F, Marin C, Aoki D, and Hanfland M 2009 *J. Phys.: Condens. Matter* **21** 232202
[23] R. Xu, A. Husmann, T. F. Rosenbaum, M.-L. Saboungi, J. E. Enderby & P. B. Littlewood, 1997 Nature **390** 57
[24] Iwasa Y, McNiff E J, Bellis R H and Sato K 1993 *Cryogenics* **33** 836
[25] Werthamer N R, Helfand E and Hohenberg P C 1966 *Phys. Rev*. **147** 295
[26] Lei Hechang, Graf D, Hu Rongwei, Ryu Hyejin, Choi E S, Tozer S W, and Petrovic C, 2012 *Phys. Rev. B* **85** 094515
[27] Nakayama K, Sato T, Richard P, Karahara T, Sekiba Y, Qian T, Chen G F, Luo J L, Wang N L, Ding H, and Takahashi T, 2010 *Phys. Rev Lett.* **105** 197001
[28] Abdel-Hafiez M, Ge J, Vasiliev A N, D. Chareev D A, Van de Vondel J, Moshchalkov V V and Silhanek A V 2013 *Phys. Rev. B* **88** 174512
[29] Blatter G, Feigel'man M V, Geshkenbein V B, Larkin A I and Vinokur V M 1994 *Rev. Mod. Phys*. **66** 1125
[30] Lei H C, Hu R W, Choi E S and Petrovic C 2010 *Phys. Rev. B* **82** 134525
[31] Yeshurun Y and Malozemoff A P 1988 *Phys. Rev. Lett*. **60** 2202

## Figure captions

**Fig. 1** XRD patterns of undoped and Ag-doped $FeSe_{0.94}$

**Fig. 2** Temperature dependence of the resistivity for the undoped and the Ag doped $FeSe_{0.94}$ samples at temperatures up to 300 K. Inset: The same data at temperatures near $T_c$

**Fig. 3** Temperature dependence of FC magnetization measured at 20 Oe and of derivative of resistivity at $H = 0$ for the undoped sample. Anomalies due to a structural phase transition and due to magnetic changes, both occurring at ~ 80K, are marked by the red line.

**Fig. 4** Comparison of the superconducting transition curves for the Ag doped $FeSe_{0.94}$ sample and a β-FeSe single crystal [9].

**Fig. 5** Temperature dependence of the resistivity at applied magnetic fields up tot 14 T: (a) – undoped molten $FeSe_{0.94}$ sample, (b) – Ag doped $FeSe_{0.94}$ sample

**Fig. 6** Field dependence of the magnetoresistance $\Delta\rho/\rho_o$ - with $\Delta\rho= (\rho(T,B) - \rho_o)$ and $\rho_o=\rho(T,0$ - for Ag-doped $FeSe_{0.94}$ at different temperatures. Inset: The same magnetoresistance data in a Kohler plot, where the magnetic field axis was normalized by the factor $\rho(300\text{ K})/\rho_o$

**Fig. 7** Temperature dependence of the upper critical field defined at the midpoint of the transition curves for both undoped and the Ag-doped samples. Red lines: Fit of the experimental $B_{c2}(T)$ data by the WHH model neglecting spin-paramagnetic effects and ignoring spin-orbit scattering

**Fig. 8** Field dependence of the magnetic moment for the Ag doped sample at 2K in the range of low magnetic fields

**Fig. 9** Arrhenius plots lnρ vs 1/T for (a) undoped $FeSe_{0.94}$ and (b)Ag-doped $FeSe_{0.94}$. The data are fitted by the straight lines according to Eq. (4)

**Fig. 10** Plots of $\ln\rho_o(H)$ vs $U_o/k_B$ for the undoped and Ag doped samples. The data are fitted by the straight lines according to Eq. (5)

**Fig. 11** Field dependence of the activation energy for undoped and Ag doped $FeSe_{0.94}$ and a β-FeSe single crystal [9] in a double-logarithmic plot. The data are fitted by $U_o \sim H^{-\alpha}$ shown as straight lines. The arrow marks the crossover from single-vortex pinning to collective pinning (see text)

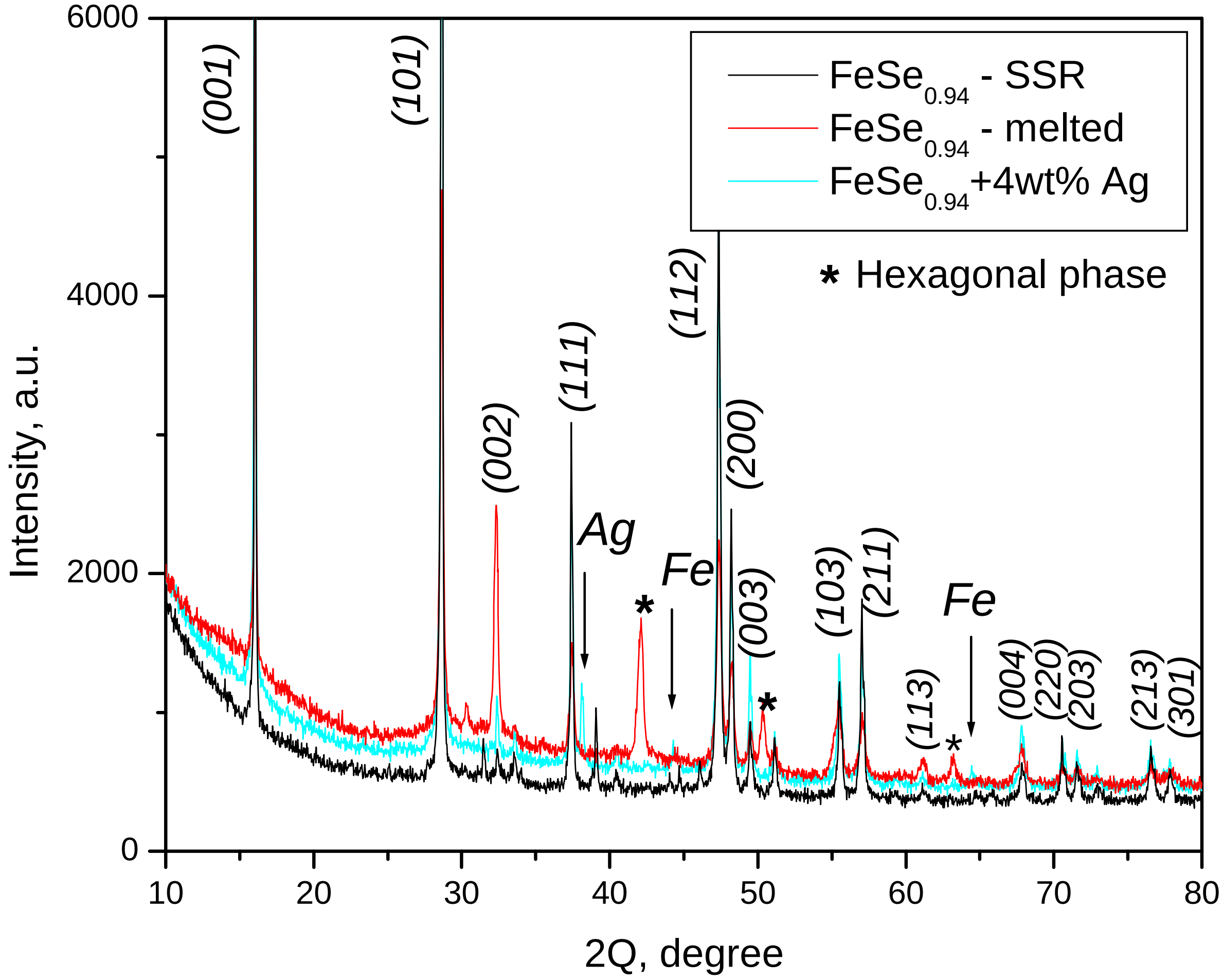

**Fig. 1**

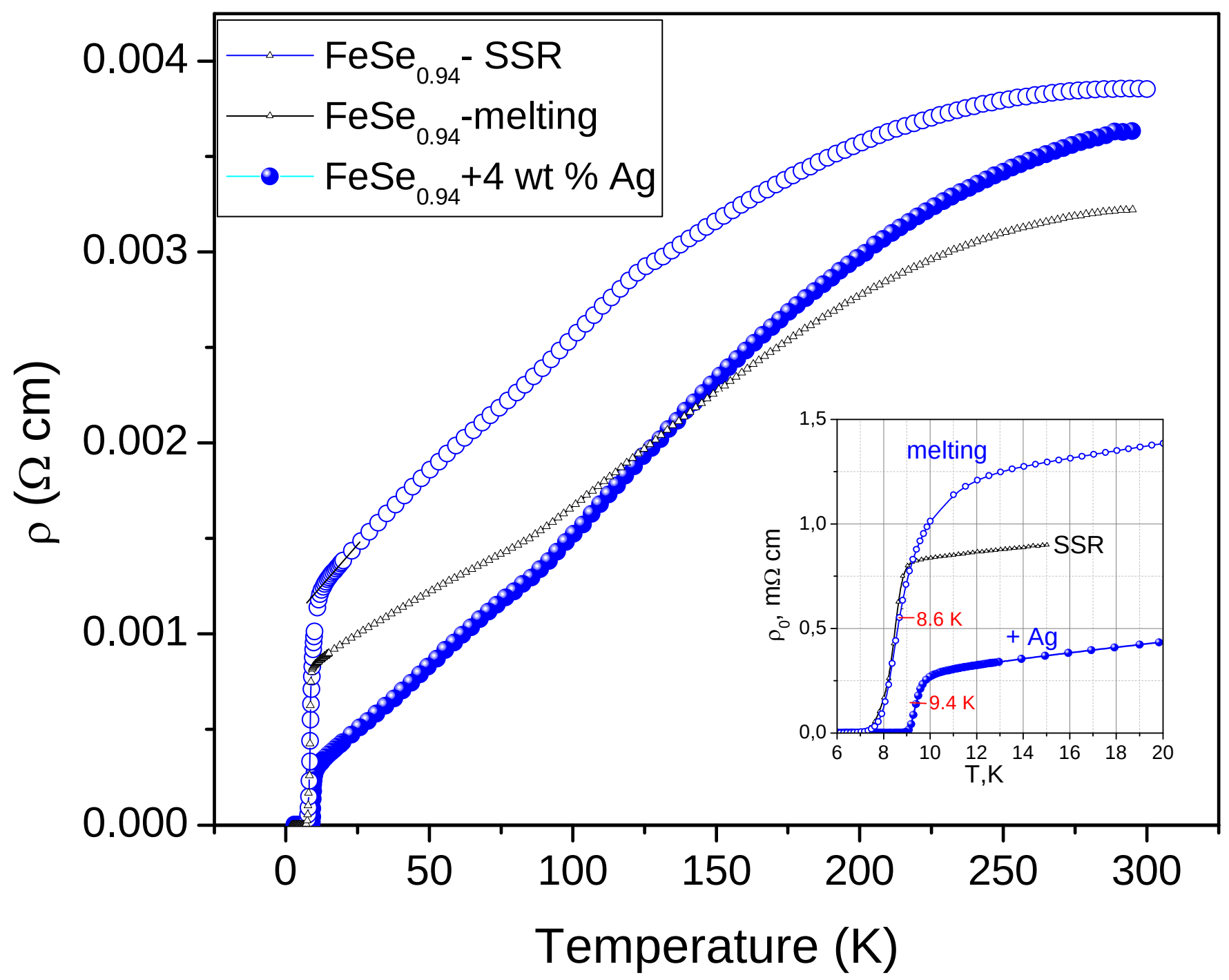

Fig. 2

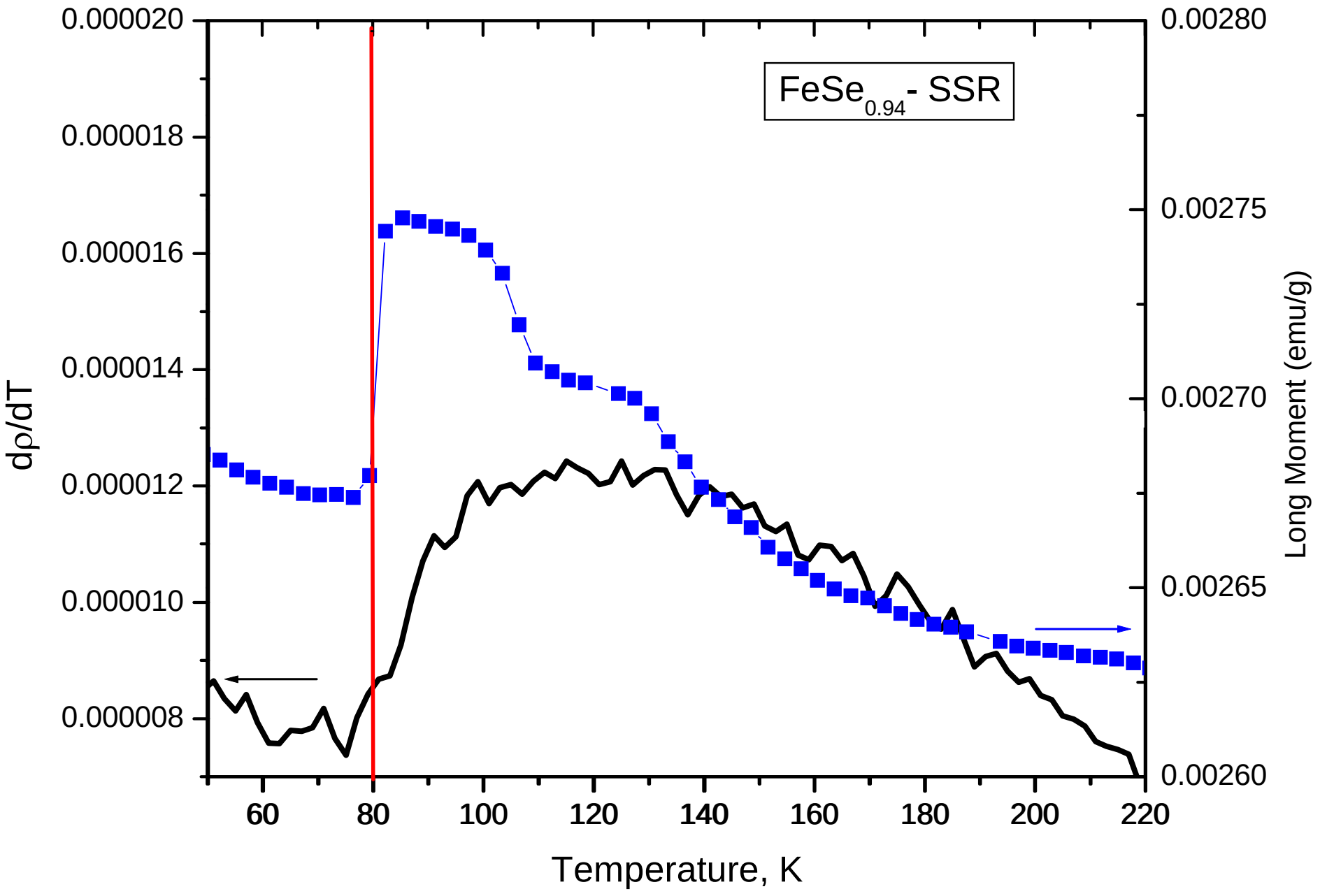

FeSe0.94- SSR
dρ/dT
Long Moment (emu/g)
Temperature, K

Fig. 3

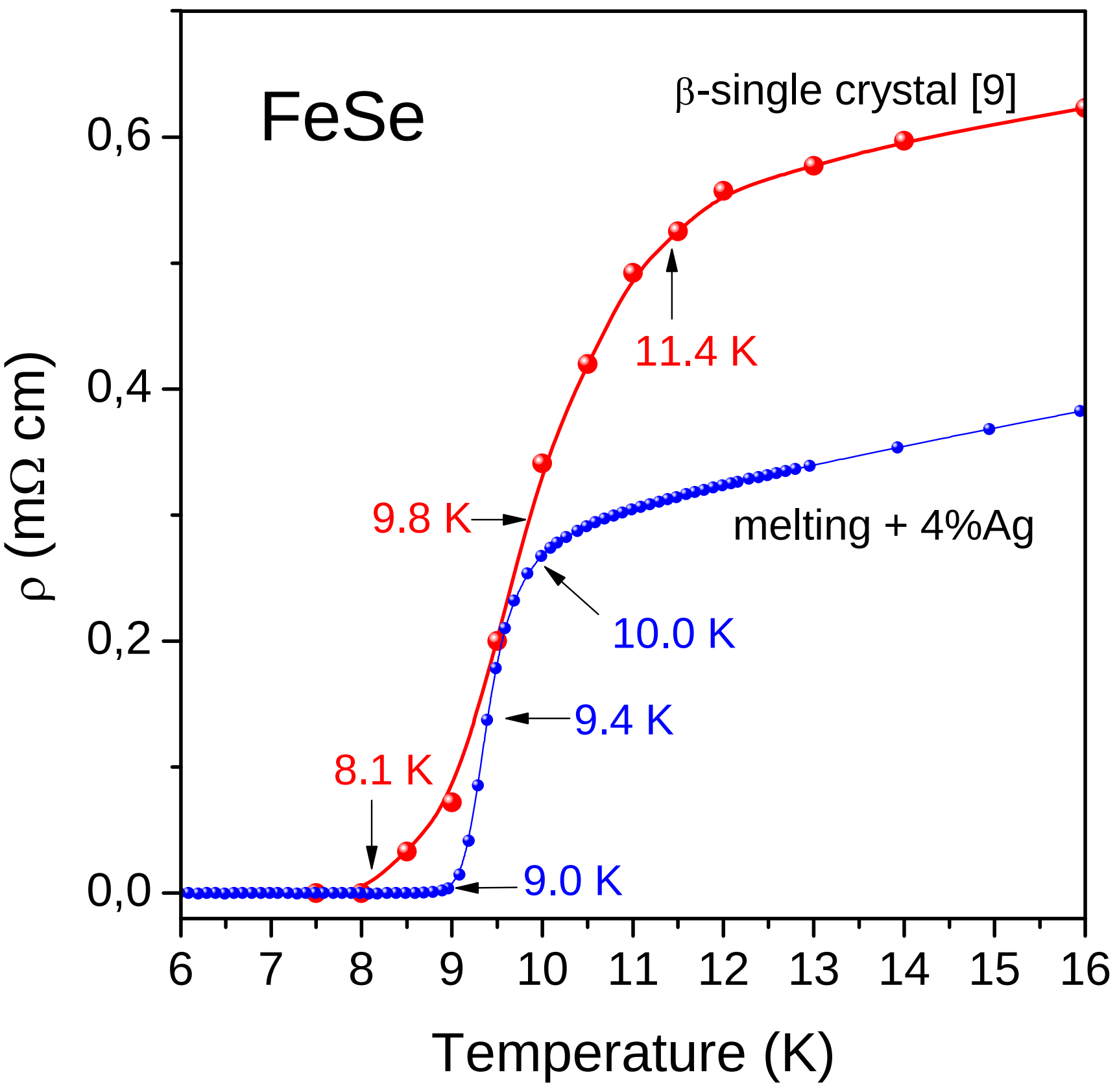

**Fig. 4**

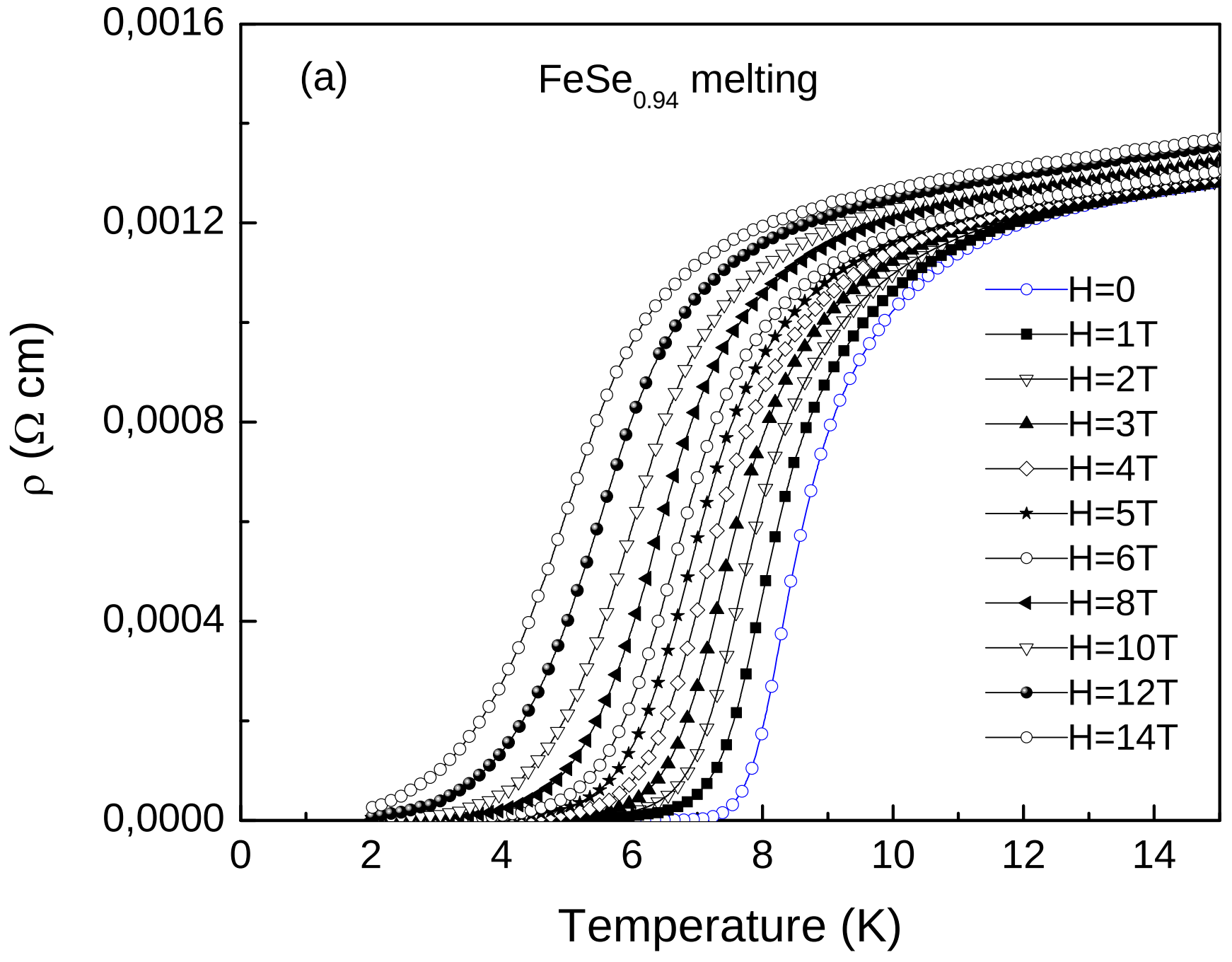

**Fig. 5a**

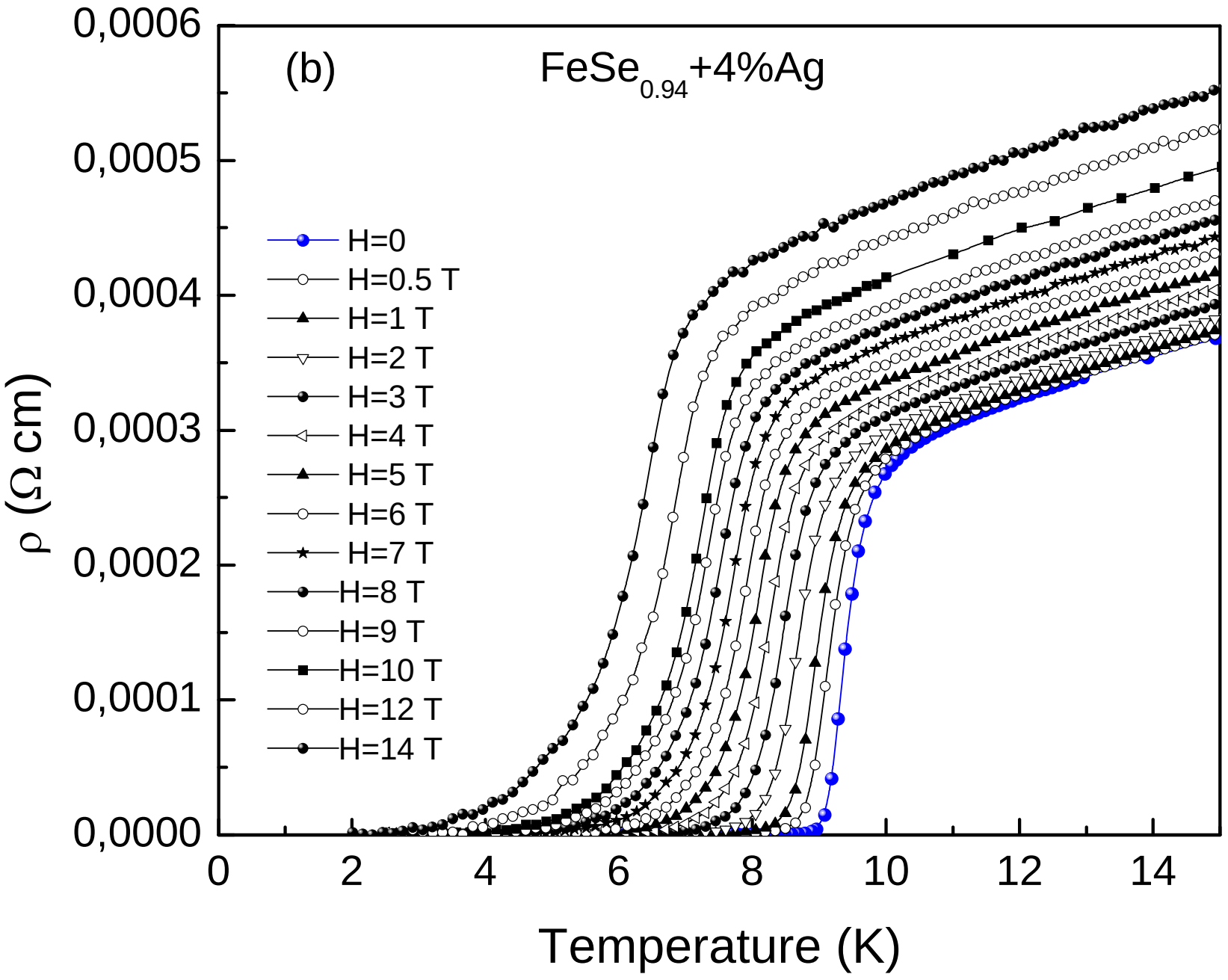

**Fig. 5b**

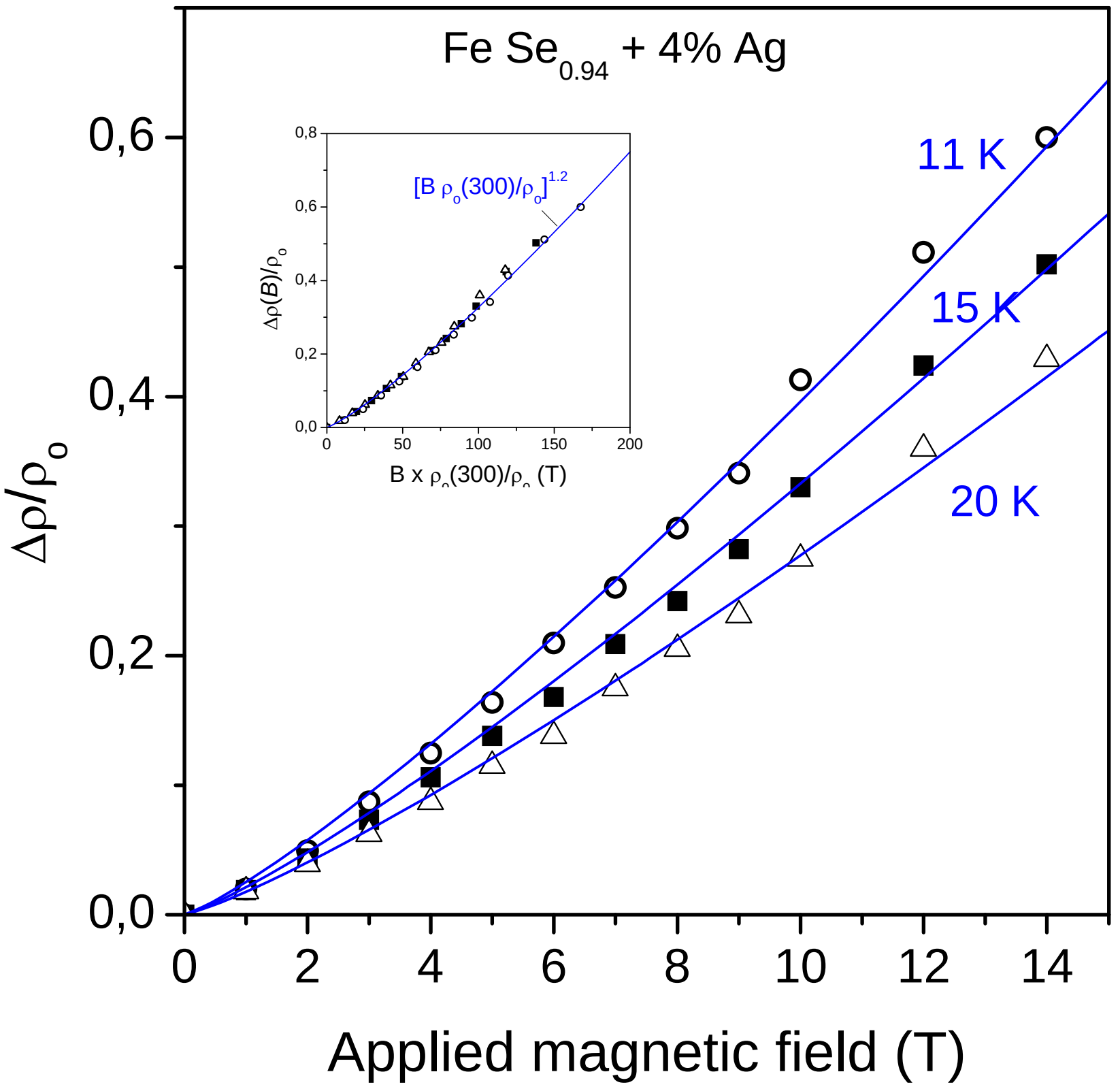

Fe Se0.94 + 4% Ag
Δρ/ρo
Applied magnetic field (T)
11 K
15 K
20 K
[B ρo(300)/ρo]1.2
Δρ(B)/ρo
B x ρo(300)/ρo (T)

**Fig. 6**

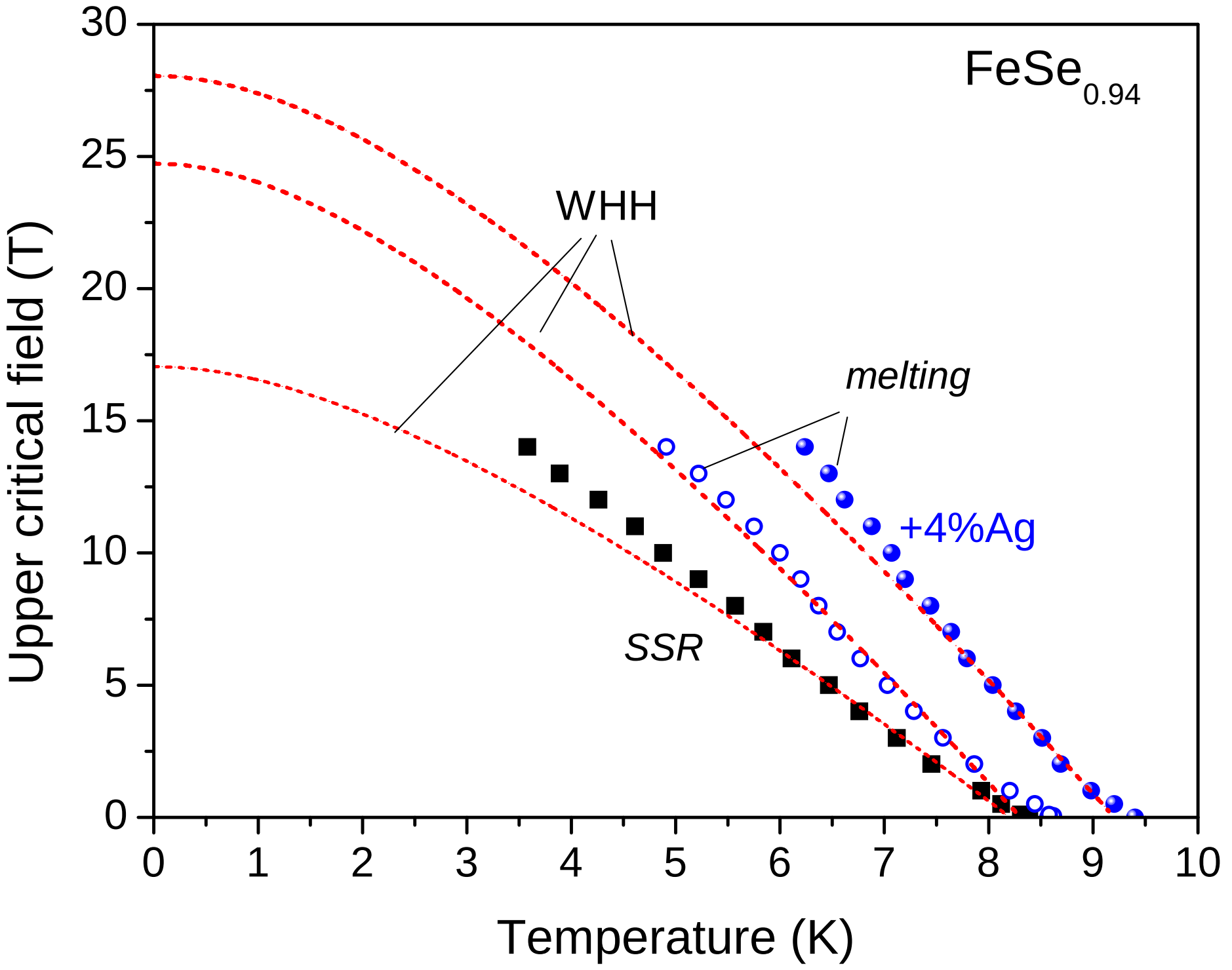

**Fig. 7**

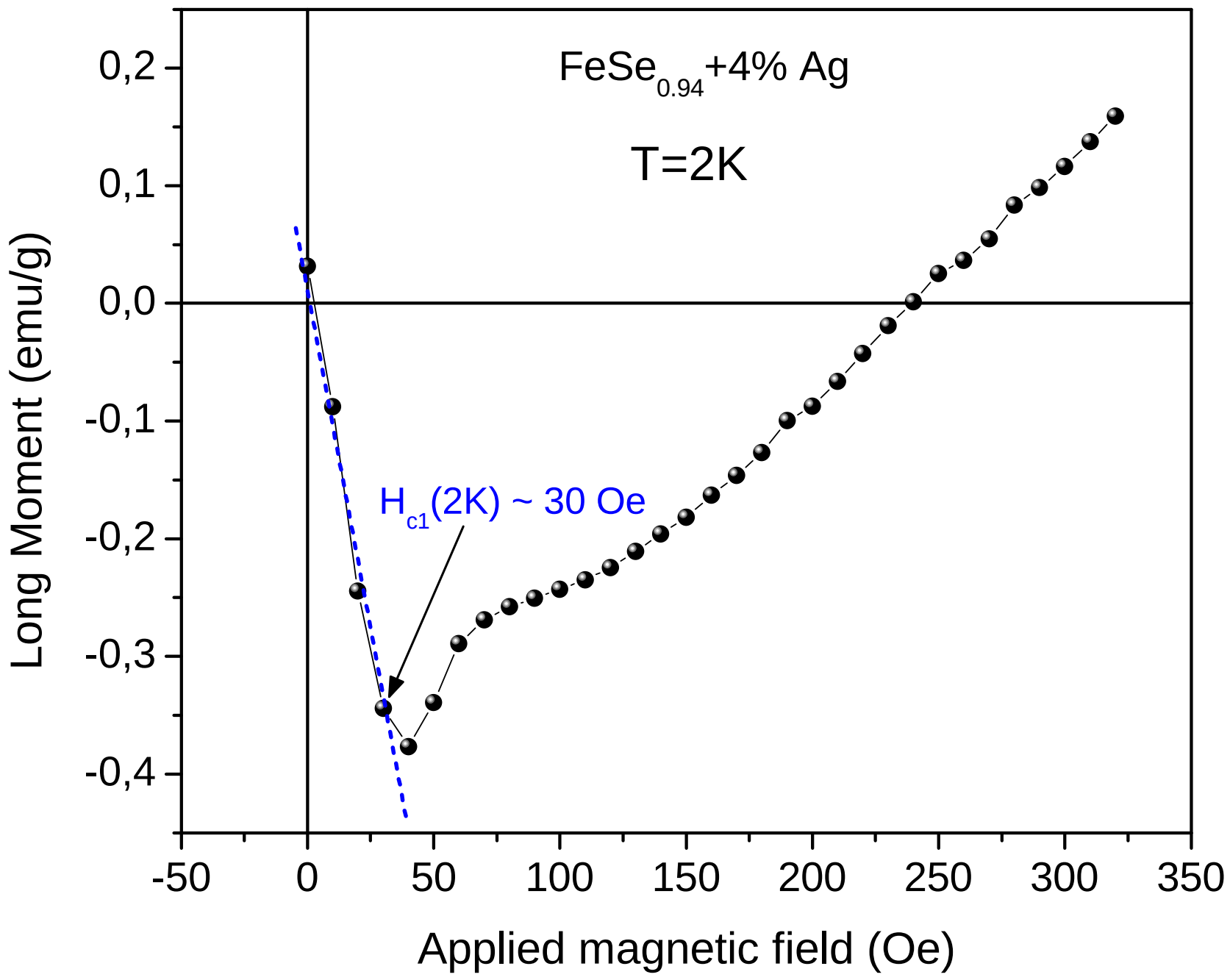

**Fig. 8**

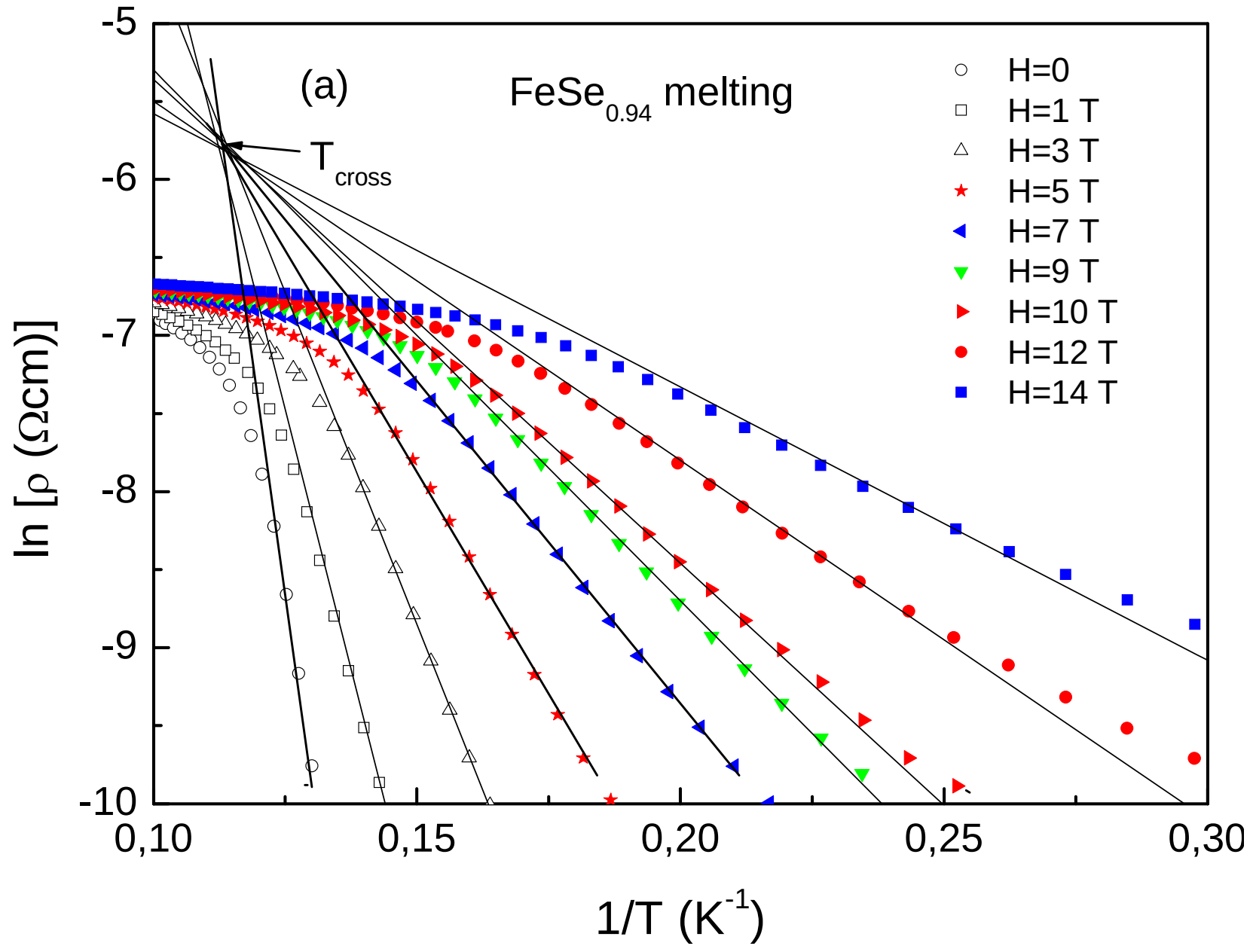

**Fig. 9a**

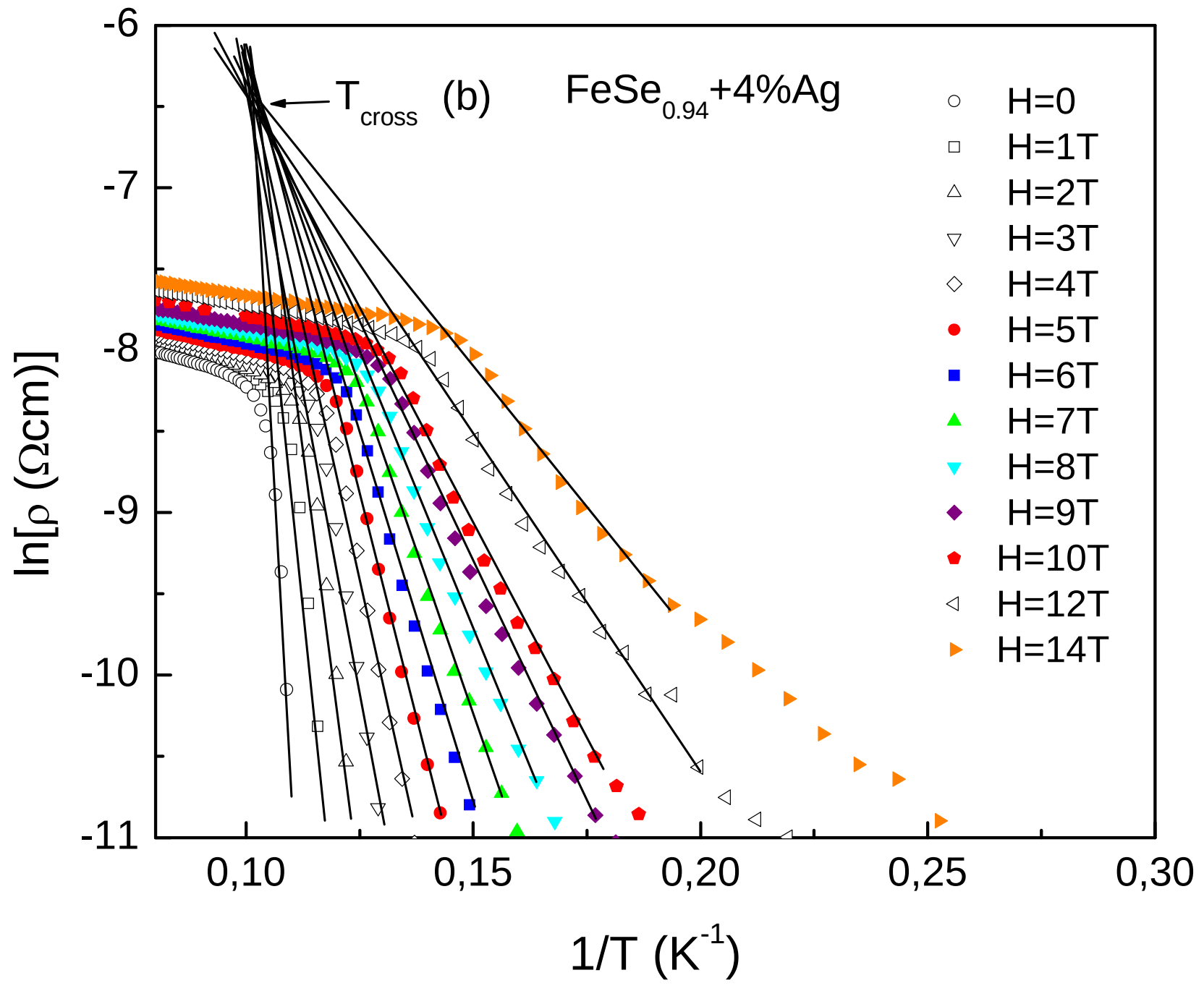

**Fig. 9b**

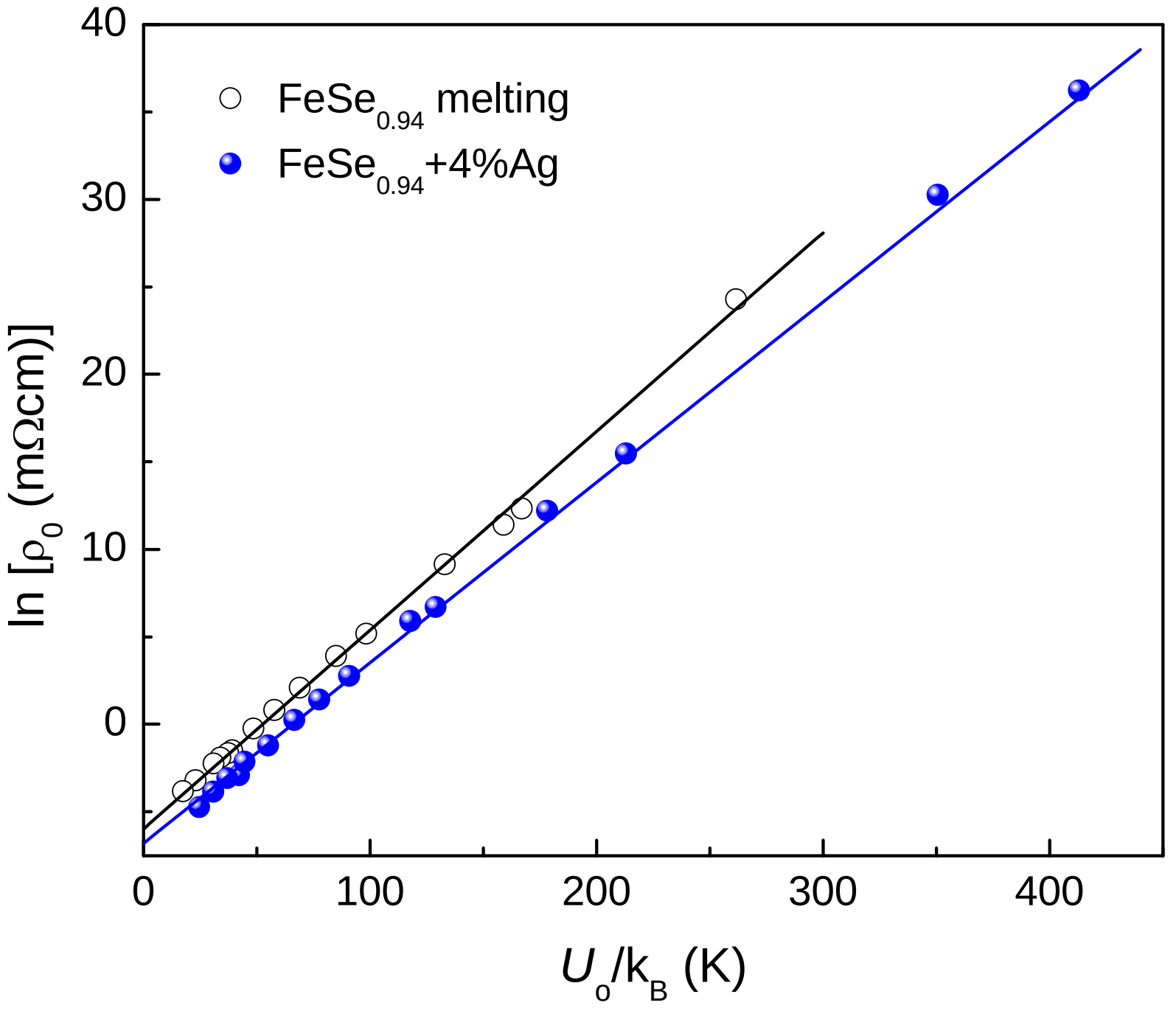

Fig. 10

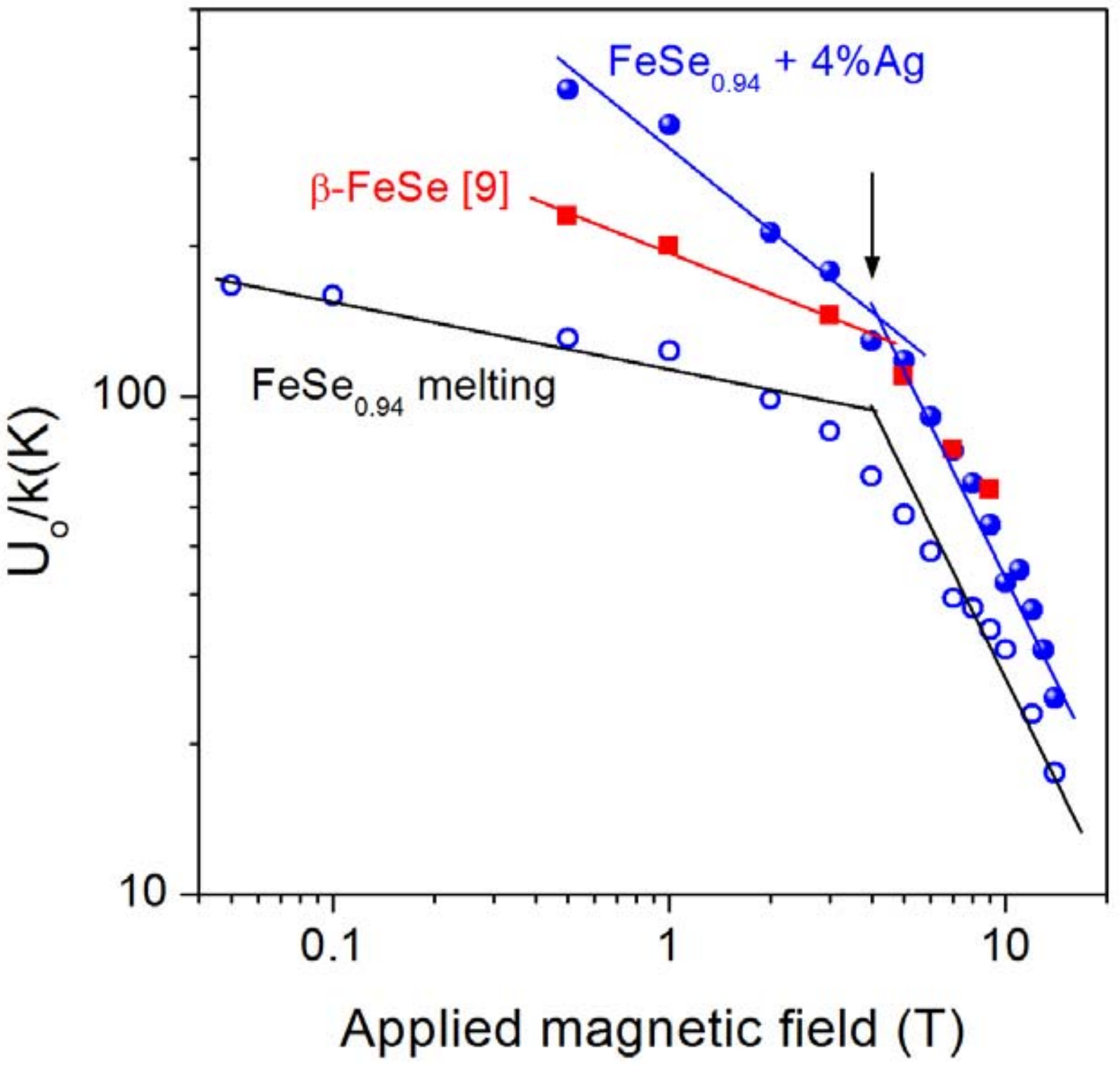

Fig. 11

**Table 1** Superconducting parameters of the investigated $FeSe_{0.94}$ samples and of a β-FeSe single crystal [9] : $T_c^{50}$ – superconducting transition temperature (midpoint of transition), RRR- residual resistivity ratio, $B_{c2}^{50}(0)$ – orbital upper critical field at $T$=0 according to the WHH model (Eq. (1)), ξ(0) - Ginzburg-Landau coherence length at $T$=0 (see Eq. (2)), $U_o/k_B$ - activation energy at 1T.

| **Sample** | $T_c^{50}$ (K) | **RRR** | $B_{c2}^{50}(0)$ (T) | $(dB_{c2}/dT)_{Tc}$ (T/K) | **ξ(0)** (nm) | $U_o/k_B$ at 1T (K) |
|---|---|---|---|---|---|---|
| $FeSe_{0.94}$ (SSR) | 8.4 | 3.2 | 17.0 | 3.0 | 4.33 | 43 |
| $FeSe_{0.94}$ melting | 8.6 | 2.94 | 24.8 | 4.3 | 3.58 | 120 |
| $FeSe_{0.94}$ melting +4% Ag | 9.4 | 9.5 | 28.1 | 4.4 | 3.37 | 351 |
| β-FeSe single crystal [9] | 9.8 | 14 | 18.0 | 2.65 | 4.28 | 200 |

**Table 2** Superconducting transition temperatures of undoped and Ag doped molten samples obtained from ρ(T) data ($T_c^{50}$), from the Arrhenius plots in Fig. 9 ($T_c^{cross}$) and from ln $\rho_o$-$U_o$ plots in Fig. 10 ($T_c^{fit}$)

| **Sample** | $T_c^{50}$ (K) | $T_c^{cross}$ (K) | $T_c^{fit}$ (K) |
|---|---|---|---|
| $FeSe_{0.94}$ melting | 8.6 | 8.85 | 8.80 |
| $FeSe_{0.94}$ melting +4% Ag | 9.4 | 9.82 | 9.70 |